\documentclass[12pt]{article}
\usepackage{sigsam, amsmath}

\usepackage[textsize=tiny]{todonotes}

\usepackage{csquotes}
\usepackage[sort,noadjust]{cite}

\usepackage{siunitx}
\usepackage{soul}
\usepackage[svgnames]{xcolor}
\usepackage{minted}

\usepackage{xspace}
\newcommand\Julia{\texttt{Julia}\xspace}

\newcommand\OSCAR{\texttt{OSCAR}\xspace}
\newcommand\mrdi{\texttt{mrdi}\xspace}

\newcommand\Macaulay{\texttt{Macaulay2}\xspace}

\newcommand\ZZ{{\mathbb Z}}
\newcommand\QQ{{\mathbb Q}}
\newcommand\PP{{\mathbb P}}

\issue{TBA}
\articlehead{TBA}
\titlehead{Interprocess Communication of Algebraic Data}
\authorhead{Antony Della Vecchia}
\begin{document}

\title{Interprocess Communication of Algebraic Data}

\author{Antony Della Vecchia \\
  TU Berlin \\
  Berlin, Germany, 10623 \\
  \url{vecchia@math.tu-berlin.de}}

\date{}

\maketitle

\begin{abstract}
  We discuss implementation details of \OSCAR's serialization framework, highlighting the design decisions that allow the fine tuning of serialization methods for specific use cases.
  In particular, we show how the \mrdi file format can be used for distributed computing.
\end{abstract}

\section{Introduction}

Serialization is the process of translating data in main memory to a format that can be stored or transmitted such that the data can be reconstructed on a different process or even machine.
Although effective methods exist for general-purpose computing, serialization of algebraic data is a difficult problem, the main obstacle being the reconstruction of the correct algebraic context, which we elaborate on in Section \ref{sec:file-format}.
The \mrdi file format \cite{DellaVecchiaJoswigLorenz:2024}, developed as part of the Mathematics Research Data Initiative (MaRDI) \cite{the_mardi_consortium_2022_6552436}, was designed to address this problem.
Although the focus of \cite{DellaVecchiaJoswigLorenz:2024} was set on long-term storage, the format can be used for broader applications such as inter-process communication for distributed computations.
Our open source implementation serves as a blueprint for other systems looking to adopt similar approaches.

\OSCAR \cite{OSCAR, OSCAR-book} is a free and open source computer algebra system written in the \Julia programming language \cite{Julia-2017}.
\OSCAR provides a core serialization framework and implementations for serializing (and deserializing) all core types to (and from) the \mrdi file format.
Building on tools for distributed computing in \Julia's standard library, \OSCAR provides support for multiprocess methods with algebraic types.
The details are discussed in Section \ref{sec:framework}.

Section \ref{sec:use-cases} presents two specific use cases demonstrating significant speedups: computing components of the vanishing ideal of the fourth secant variety of $\PP^3 \times \PP^3 \times \PP^3$ up to degree five \cite{boege2026}, and computing the determinant of a matrix with entries in a univariate polynomial ring over $\ZZ$. The example code is available at \url{https://github.com/dmg-lab/OscarParallelExamples} \enspace .

\section*{Acknowledgements}
  We thank Matthias Zach and Friedemann Groh for providing the matrix, and Jereon Hanselman for many helpful discussions about the format.
  This work was the Deutsche Forschungsgemeinschaft (DFG), project number 460135501, NFDI 29/1 \enquote{MaRDI – Mathematische Forschungsdateninitiative}.

\section{The \mrdi File Format}
\label{sec:file-format}
The \mrdi file format is designed for algebraic context reconstruction. For example, two polynomials can only be multiplied if the computer algebra system recognizes them as elements of the same polynomial ring, where recognition means their parent rings point to the same memory location<.
The key feature of the format is the ability to transfer this context between different \OSCAR sessions and versions, with plans within the MaRDI project to support transfers between different computer algebra systems; see \url{https://github.com/JHanselman/MagmaMaRDI-JSON} for a  Magma \cite{Magma} prototype.
The format is JSON-based and, similar to the \texttt{polymake} file format \cite{polymake_XML:ICMS_2016} and has the structure of an annotated tree with four main subtrees: \texttt{\_type}, \texttt{data}, \texttt{\_refs}, and \texttt{\_ns}. The \texttt{\_ns} subtree records which system and version serialized the file, \texttt{\_type} indicates the type of the stored object and its parameters, \texttt{\_refs} stores context objects identified by UUIDs, and \texttt{data} stores the actual serialized data.
See Figure \ref{fig:p-example} for an example.
JSON was chosen for its human readability: files can be inspected without specialized tools, upgrade scripts can operate via pure text manipulation, and the format is free to evolve without requiring a fixed optimal representation from the outset. Such an approach has been successfully employed with the \texttt{polymake} format for over twenty years.

\begin{figure}[h]
\begin{minted}{js}
{ "_ns": {"Oscar": [ "https://github.com/oscar-system/Oscar.jl", "1.8.0" ] },
  "_type": { "name": "MPolyRingElem",
	     "params": "c33bb5fb-96d7-42ed-8cd9-89b83eb7a298"},
  "data": [ [ [ 3, 0 ], "1" ], [ [ 1, 1 ], "-1" ], [ [ 0, 0 ], "1" ] ],
  "_refs": {
    "c33bb5fb-96d7-42ed-8cd9-89b83eb7a298": {
      "_type": { "name": "MPolyRing", "params": {"_type": "QQField" } },
      "data": {"symbols": [ "x", "y" ] } } } }
\end{minted}
  \caption{%
    \label{fig:p-example}
    The bivariate polynomial $p = x^3 - xy + 1$ with coefficients $\QQ$ in the \mrdi file format.
  }
\end{figure}

\section{\OSCAR's Serialization Framework}
\label{sec:framework}
We first describe how \OSCAR's serialization framework operates for long-term storage, by illustrating the process using the bivariate polynomial $p = x^3 -xy + 1 $ over $\QQ$ shown in Figure \ref{fig:p-example}.
Serialization proceeds as follows: first the \texttt{\_ns} is written, then a shallow pass through $p$ builds the \texttt{\_type} subtree, determining the type \texttt{MPolyRingElem} (element of a mulitvariate polynomial ring) and registering the parent ring with a UUID in the \texttt{SerializerState}, and finally the terms of $p$ are written to \texttt{data} as an array of tuples representing exponent vectors and coefficients.
All registered parameters are then written to \texttt{\_refs}, in this case the parent ring encoded with its coefficient ring and generator symbols. Deserialization reverses this process: \OSCAR reads the \texttt{\_type} subtree, checks whether the parent ring has already been deserialized via its UUID, deserializes it from \texttt{\_refs} if not, and reconstructs $p$ in the correct algebraic context. A \texttt{GlobalSerializerState} allows contexts to be shared across multiple files, a feature that is also necessary for multi-process methods.

While the above suffices for long-term storage, the resulting files are quite verbose.
For inter-process communication, \OSCAR uses its \texttt{IPCSerializer}, omitting the \texttt{\_ns} subtree, as each process runs the same version of \OSCAR, and the \texttt{\_refs} subtree by ensuring contexts are preloaded on each receiving process.
To manage this automatically, \OSCAR provides an \texttt{OscarWorkerPool}, a subtype of \Julia's \texttt{AbstractWorkerPool}, which handles context distribution via a post-order traversal of the \texttt{\_type} subtree, guaranteeing each context is sent before the types that depend on it.
This allows users to use functions from \texttt{Distributed.jl} such as \texttt{remotecall} and \texttt{pmap} directly with \OSCAR types.
Further optimizations modify the \texttt{data} subtree and rely on \Julia's parametric type system and multiple dispatch: the \texttt{SerializerState} and \texttt{DeserializerState} are parameterized by the serializer type, allowing dispatch to select a more efficient encoding over the default when one exists.
Figure \ref{fig:save-snippet} shows multiple encodings for parts of the \texttt{data} subtree that involve univariate polynomials.
For long-term storage the encoding is sparse and avoids ambiguities in the order of coefficients.

\begin{figure}[h]
\begin{minted}{jl}
function save_object(s::SerializerState, p::PolyRingElem)
  save_object(s, [
    (i - 1, c) for (i, c) in enumerate(coefficients(p))
      if !is_zero(c) ])
end

function save_object(s::SerializerState{IPCSerializer}, p::PolyRingElem)
  save_object(s, collect(coefficients(p)))
end
\end{minted}
  \caption{%
    \label{fig:save-snippet}
    A snippet with the two encodings for univariate polynomials. 
  }
\end{figure}

\section{Use Cases}
\label{sec:use-cases}
We note that small computations are unlikely to benefit from these methods, as the overhead of inter-process communication outweighs any potential gain.
Determining which problems are large enough to warrant parallelization is still an active area of experimentation.

\paragraph{Computing Ideal Components from Algebraic Statistics.}
The main algorithm in \cite{CummingsHollering26} computes the kernel of a map $\phi$ homogeneous with respect to a $\ZZ^k$-multigrading up to a given total degree by iterating over multidegrees and constructing matrices whose rows are indexed by monomials of that multidegree.
Row reducing these matrices over $\QQ$ determines the kernel components degree by degree, avoiding a full elimination algorithm.
The original \Macaulay implementation can compute examples with roughly 50,000 monomials and 25,000 multidegrees, but won't terminate on an example with 2,829,056 monomials over 637,440 multidegrees.
The \OSCAR implementation \texttt{components\_of\_kernel} developed as a part of \cite{boege2026} can handle this example, however the example is still too small for parallel methods,
see \cite{boege2026} for a time comparison between the two implementations.
When examples are large enough to warrant the overhead, the row reduction steps may be handle independently by seperate processes.
The authors of \cite{boege2026} were able to compute the components of the vanishing ideal of the fourth secant variety of the Segre variety $\PP^3 \times \PP^3 \times \PP^3$ up to total degree five an example with 8,303,632 monomials over 175,616 multigrees in under three hours using an Intel Xeon Silver 4216 with 32 processes and \SI{150}{\giga\byte} of main memory, an example that would take \OSCAR over 24 hours if computed with a single process. 

\paragraph{Modular Determinant Computations for Real Plane Algebraic Curves.}
Computing discriminants of curves arising from recent work on smooth real plane algebraic curves of degree seven \cite{geiselmann2026121patchworkedcurvesdegree} is of interest.
Following an approach communicated privately by Bernd Sturmfels, Matthias Zach and Friedemann Groh provided a matrix whose determinant realizes one such discriminant.
The matrix $M$ is of size $48 \times 48$ with entries in a univariate polynomial ring over $\ZZ$.
A standard approach \cite{MR3087522} is to reduce the matrix over sufficiently many primes and then use the Chinese Remainder Theorem to find the determinant with coefficients in $\ZZ$.
This approach computes the determinant of $M$ in 1326 seconds on an Intel Core i7-3930K with \SI{150}{\giga\byte} of main memory using a single process and in 357 seconds using six processes.

\bibliographystyle{acm}
\bibliography{biblio.bib}

\end{document}